\documentclass{elsarticle}
\usepackage{graphicx}
\usepackage{dcolumn}% Align table columns on decimal point
\usepackage{bm}% bold math
\usepackage{rotating}
\usepackage{natbib}
\usepackage{txfonts}
\def\be{\begin{equation}}
\def\ee{\end{equation}}
\def\bg{\begin{eqnarray}}
\def\en{\end{eqnarray}}
\def\nn{\nonumber}

\renewcommand{\vec}[1]{\boldsymbol{#1}}

\newcommand{\La}{\ensuremath{\mathcal L}}

\begin{document}

\begin{frontmatter}
%\preprint{ADP-12-10/T777}

\title{Structure and Coulomb dissociation of $^{23}$O within the quark-meson 
coupling model}

\author[ROORKEE]{R. Chatterjee}
\author[SAHA]{R. Shyam}
\author[CSSM,NATAL]{K. Tsushima}
\author[CSSM1]{A. W. Thomas}

\address[ROORKEE]{Physics Department, Indian Institute of Technology, 
Roorkee, India} 
\address[SAHA]{Saha Institute of Nuclear Physics, 1/AF Bidhan Nagar, Kolkata, India}
\address[CSSM]{Special Research Centre for the Subatomic Structure of Matter (CSSM),
School of Chemistry and Physics, University of Adelaide, SA 5005, Australia}
\address[NATAL]{International Institute of Physics, Federal University of
Rio Grande do Norte, Natal, Brazil}
\address[CSSM1]{Special Research Centre for the Subatomic Structure of Matter (CSSM) 
and ARC Centre of Excellence in Particle Physics at Terascale (CoEPP), School of 
Chemistry and Physics, The University of Adelaide, SA 5005, Australia}

\begin{abstract}

We study the ground-state structure of nuclei in the vicinity of the one-neutron
dripline within the latest version of the quark-meson coupling (QMC) model with a 
particular emphasis on $^{23}$O. For this nucleus the model predicts a 
$^{22}$O(0$^+$)$\otimes$\,$n$2$s_{1/2}$ configuration for its ground state, with a 
one neutron separation energy in close agreement with the corresponding experimental 
value. The wave function describing the valence neutron-core relative motion was then 
used to calculate the Coulomb dissociation of $^{23}$O on a lead target at a beam 
energy of 422 MeV/nucleon. The experimental neutron-core relative energy spectrum and 
the total one-neutron removal cross sections are well described by the calculations. 
The widths of the longitudinal momentum distributions of the $^{22}$O fragment are 
found to be broad, which do not support the formation of a neutron halo in this 
nucleus.

\end{abstract}


\end{frontmatter}

\section{Introduction}

In recent years a lot of effort has been devoted to study the structure of the light 
neutron rich nuclei that have some unusual properties. For example, the neutron drip-line 
for the oxygen isotopes is located at $^{24}$O (see e.g. Refs.~\cite{gui90}), and not at 
$^{28}$O, the doubly magic nucleus that has been found to be unbound. This suggests that 
a new magic number appears at $N=16$ in neutron rich nuclei. The reason for this 
observation has been investigated in several studies~\cite{ots01,jia05}. Otsuka et al.
\cite{ots01} have suggested  that it should be attributed to the spin-isospin dependent 
part of the nucleon-nucleon force yielding a strongly attractive interaction between a 
proton in $1d_{5/2}$ orbit and a neutron in $1d_{3/2}$ orbit, which in turn increases the 
gap between $2s_{1/2}$ and $1d_{3/2}$ orbitals for the $N$=16 oxygen isotope.

As the neutron drip-line is approached the nuclei experience weakening of the neutron 
binding energies, which leads to some special effects.  The sudden rise of the interaction 
cross sections found in Refs.~\cite{tan85} for several lighter isotopes closer to the 
neutron drip-line was attributed to the extended density distribution(s) of the valence
neutron(s) usually referred to as a neutron halo~\cite{han87}. For oxygen isotopes a 
sudden rise of about 12$\%$ was observed~\cite{kan01} in the interaction cross sections 
between $^{22}$O and $^{23}$O. This led these authors to suggest $^{23}$O as a 
candidate for a neutron-halo system. However, there are several other observations that 
are not consistent with this picture. The valence neutron in $^{23}$O is relatively 
strongly bound, with a one neutron separation energy ($S_{-n}$) of 2.74 MeV, which is in
contrast to the well established one neutron halo nuclei like, $^{11}$Be and $^{19}$C, 
that have $S_{-n}$ of only about 0.5 MeV. Thus the last neutron in $^{23}$O should not 
extend to larger distances. Furthermore, the longitudinal momentum ($p_\|$) (with 
respect to the reaction plane) distributions (LMD) of the $^{22}$O fragment have been 
measured~\cite{kan02,sau04,gil04,taj10} in the breakup reactions of $^{23}$O on 
lighter target nuclei, where the widths of these distributions are found to be in 
excess of 100 MeV/c. This is about 2-3 times larger than those observed in the cases of 
$^{11}$Be and $^{19}$C induced reactions (see, e.g. Ref.~\cite{jon04}). A narrow LMD is a 
direct signature of the development of the halo structure (see, e.g. the discussions 
presented in~\cite{ber93}). 

Moreover, there has been a disagreement in the assignment of the ground state spin-parity 
($J^\pi$) of $^{23}$O nucleus among various authors. In Ref.~\cite{kan02}, from the LMD 
measurements of the $^{22}$O and $^{21}$O fragments in the $^{23}$O induced reaction on 
a carbon target, it was concluded that the data support a $J^\pi$ of $\frac{5}{2}^+$. 
However, this conclusion was not supported by the measurements of the $^{23}$O breakup 
reactions reported in Ref.~\cite{gil04,noc05} where a $J^\pi$ of $\frac{1}{2}^+$ was 
found to be consistent with the data. In Ref.~\cite{kan11} the interaction cross sections 
of $^{22}$O and $^{23}$O isotopes were measured on a carbon target and it was concluded 
that these data are more in agreement with a $^{22}$O + valence neutron in the $2s_{1/2}$ 
picture of the $^{23}$O nucleus. This would imply a $J^\pi$ of $\frac{1}{2}^+$. These 
conflicting experimental results call for a more rigorous theoretical investigation of 
the $^{23}$O ground state configuration. Furthermore, there is a need to analyze the 
data on the breakup reactions of $^{23}$O within a more microscopic theoretical model 
than has been employed so far. 

The aim of this paper is to investigate within the quark-meson coupling (QMC) model, 
the ground state structure of several light neutron rich nuclei that lie in the vicinity 
of the one-neutron dripline with a particular attention to the $^{23}$O nucleus. The
QMC model is a quark-based model for nuclear matter, finite nuclei and hypernuclei
\cite{gui88,gui96,sai96,tsu98,sai07,gui08}, where quarks in the non-overlapping MIT bags 
interact self consistently with isoscalar-scalar ($\sigma$) and isoscalar-vector 
($\omega$) mesons in the mean field approximation. The explicit treatment of the nucleon 
internal structure is a key point of this model and it represents an important departure 
from quantum hadrodynamics (QHD)~\cite{ser86}. The self-consistent response of the bound 
light quarks to the mean $\sigma$ field leads to a new saturation mechanism for nuclear 
matter~\cite{gui88}. 
 
Although formulated as a relativistic mean field theory with just a few parameters 
(determined from the properties of nuclear matter), QMC model has been shown to lead to a 
remarkably realistic Skyrme force \cite{gui04,gui06}. This has led to its application to 
describe rather successfully several properties (e.g. binding energies per particle, density 
and charge distributions and energy levels) of closed shell nuclei \cite{sai07,gui06} 
with masses spanning a wide range of the periodic table. The agreement with the 
corresponding experimental data is not been inferior to that obtained within the 
Skyrme Hartree-Fock (HF) or relativistic mean field (RMF) models.  

The same effective interaction (with identical parameters) has also been used to describe 
the properties of nuclei far from stability in Ref.~\cite{gui06} where this model was used 
to predict the positions of the two-neutron drip line in Ni and Zr isotopes. In this study 
the pairing correlation between two neutrons has been treated in a Hatree-Fock-Bogoliubov
(HFB) approach. In these calculations the neutron drip line appears around the neutron 
numbers that are similar to the predictions provided by the Skyrme force SLy4 commonly 
used in the non-relativistic calculations~\cite{cha98}. Furthermore, the shell quenching, 
which has important consequences for the astrophysical rapid capture process, is also very 
close to that obtained in the Skyrme-Hartree-Fock-Bogoliubov approach in Ref.~\cite{cha98}. 
It is remarkable because, unlike the non-relativistic Skyrme force based calculations where 
the experimental values for the binding energy and the radii are included in the fitting 
procedure for determining the corresponding parameters, in the QMC calculations the 
parameters remain the same and these quantities are actually predicted by the model. 
Therefore, the QMC model contains features that make it appealing to apply to  the 
description of the structure of the drip line nuclei.
 
In this paper we report the results of the QMC model calculations for the light neutron 
rich nuclei that lie in the vicinity of one-neutron dripline - a region which has not been 
explored within this model so far. We use the latest version of the QMC model (to be 
referred to as QMC-III)~\cite{gui08} to predict the one-neutron separation energy and the 
valence neutron spin-parity in the ground state of the neutron rich nuclei $^{23}$N, 
$^{23}$O, $^{31}$Ne, $^{35}$Mg, $^{37}$Na, $^{45}$S. Our particular attention will be 
focused on the $^{23}$O nucleus where we use this model to calculate also the neutron and 
proton density distributions in order to investigate the possible existence of a halo 
structure in this nucleus. Furthermore, we use the predicted valence neutron configuration 
in the ground state of $^{23}$O and the corresponding valence neutron-core wave function 
to investigate its Coulomb dissociation (CD) on a Pb target, for which some data exists
\cite{noc05} on the valence neutron-core relative energy differential and the total 
one-neutron removal cross sections.

In the next section, our formalism is presented where some important features of the 
QMC model are discussed. In this section we also presented a short review of the 
CD model that has been used to calculate the valence neutron-core relative energy 
differential and the total one-neutron removal cross sections for the breakup of 
$^{23}$O on a Pb target. Our results are presented and discussed in section 3. A summary 
and conclusions of our work is presented in section 4.

\section{Formalism}

\subsection{Quark-meson coupling model}

The QMC-III model includes the self-consistent effect of the mean scalar field on 
the familiar one-gluon exchange hyperfine interaction~\cite{rik07} that in free space 
leads to the $N-\Delta$ and $\Sigma-\Lambda$ mass splitting. With this development, 
QMC model has been able to explain the properties of $\Lambda$ hypernuclei for the 
$s$-state rather well, although the $p$- and $d$-states tend to be underbound. It 
also maintains the very natural explanation of the small spin-orbit force in the 
$\Lambda$-nucleus interaction that was found in an earlier version of the QMC model (to 
be referred to as QMC-I)~\cite{sai96,tsu98,tsu97}. In QMC-III, while the quality of 
results for $\Lambda$ and $\Xi$ hypernuclei is comparable to that obtained in QMC-I
\cite{tsu98}, no bound states for the $\Sigma$ states~\cite{gui08} were found in medium 
and heavy mass nuclei. This finding, which is a consequence of the extra repulsion 
associated with the increased one-gluon-exchange hyperfine interaction in medium, is in 
agreement with the non-observation of such states experimentally. The QMC-III model has 
recently been used~\cite{shy12} to study the production of $\Xi^-$ hypernuclei via the 
($K^-,K^+$) reaction, which is currently of great interest at the J-PARC facility in 
Japan.  

In order to calculate the properties of the finite nuclei, we construct a simple, 
relativistic shell model with self-consistent scalar and vector mean fields. The 
Lagrangian density for a nuclear system in the QMC model is written as~\cite{sai96}:
\begin{eqnarray}
&& {\La}_{QMC} =  \bar{\psi}_N(\vec{r})
[ i \gamma \cdot \partial - M_N(\sigma) - (\, g_\omega \omega(\vec{r})
+ g_\rho \frac{\tau^N_3}{2}b(\vec{r})
+ \frac{e}{2} (1+\tau^N_3) A(\vec{r}) \,) \gamma_0 ] \psi_N(\vec{r}) \nn\\
&&\hspace{5ex} - \frac{1}{2}[ (\nabla \sigma(\vec{r}))^2 + m_{\sigma}^2
      \sigma(\vec{r})^2 ]
+ \frac{1}{2}[ (\nabla \omega(\vec{r}))^2 + m_{\omega}^2 \omega(\vec{r})^2 ]
\nn \\
&&\hspace{5ex}+ \frac{1}{2}[ (\nabla b(\vec{r}))^2 + m_{\rho}^2 b(\vec{r})^2 ]
+\frac{1}{2} (\nabla A(\vec{r}))^2 \label{Lag2}.
\end{eqnarray}
Here $\psi_N(\vec{r})$, $b(\vec{r})$, $\omega(\vec{r})$ and $A(\vec{r})$
are, respectively, the nucleon, the  $\rho$ meson, the $\omega$ meson and 
Coulomb fields, while $m_\sigma$, $m_\omega$ and $m_{\rho}$ are the masses 
of the $\sigma$, $\omega$ and $\rho$ mesons.  $g_\omega$ and $g_{\rho}$ are 
the $\omega$-N and $\rho$-N coupling constants that are related to the 
corresponding (u,d)-quark-$\omega$, $g_\omega^q$, and $(u,d)$-quark-$\rho$, 
$g_\rho^q$, coupling constants as $g_\omega = 3 g_\omega^q$ and 
$g_\rho = g_\rho^q$, and $e$ is the proton charge.

The following equations of motion are obtained from the Lagrangian density 
Eqs.~(\ref{Lag2}):
%%%%
\begin{eqnarray}
&&[i\gamma \cdot \partial -M_N(\sigma)-
%(\, g_\omega \omega(\vec{r}) + g_\rho \frac{\tau^N_3}{2} b(\vec{r}) \nn \\
%&&\hspace{-1.4cm} + \frac{e}{2} (1+\tau^N_3)
(\, g_\omega \omega(\vec{r}) + g_\rho \frac{\tau^N_3}{2} b(\vec{r})
+ \frac{e}{2} (1+\tau^N_3)
A(\vec{r}) \,) \gamma_0 ] \psi_N(\vec{r}) =  0,
\label{eqdiracn1}
\end{eqnarray}
\begin{eqnarray}
%&&\hspace{-1.4cm}(-\nabla^2_r+m^2_\sigma)\sigma(\vec{r})  = \nn \\
%&&\hspace{-1.4cm} g_\sigma C_N(\sigma) \rho_s(\vec{r})
&&(-\nabla^2_r+m^2_\sigma)\sigma(\vec{r})  =
g_\sigma C_N(\sigma) \rho_s(\vec{r}),
\label{eqsigma}
\end{eqnarray}
\begin{eqnarray}
&&\hspace{-1.4cm}(-\nabla^2_r+m^2_\rho) b(\vec{r})  =
\frac{g_\rho}{2}\rho_3(\vec{r}),
\label{eqrho}
\end{eqnarray}
\begin{eqnarray}
&&\hspace{-1.4cm}(-\nabla^2_r) A(\vec{r})  =
e \rho_p(\vec{r}),
\label{eqcoulomb}
\end{eqnarray}
%%%
where, $\rho_s(\vec{r})$, $\rho_B(\vec{r})=\rho_p(\vec{r})+\rho_n(\vec{r})$,
$\rho_3(\vec{r})=\rho_p(\vec{r})-\rho_n(\vec{r})$ and $\rho_p(\vec{r})$ 
($\rho_n(\vec{r})$) are the scalar, baryon, third component of isovector, 
and proton (neutron) densities~\cite{tsu98}. On the right hand side of 
Eq.~(\ref{eqsigma}), a new and characteristic feature of QMC appears, 
arising from the internal structure of the nucleon, namely, 
$g_\sigma C_N(\sigma)= - \frac{\partial M_N(\sigma)} {\partial \sigma}$, 
where $g_\sigma \equiv g_\sigma (\sigma=0)$. We use the density dependent 
nucleon mass $M_N(\sigma)$ as parameterized in Ref.~\cite{gui08}. 

The coupled non-linear differential Eqs.~(2)-(5) can be solved by a 
standard iteration procedure as discussed in, e.g., Refs.~\cite{hor81}.
The coupling constants $g_\sigma$, $g_\omega$, $g_\rho$, and masses 
$m_\sigma$, $m_\omega$ and $m_\rho$ have been taken to be the same as those
given in Ref.~\cite{gui08}. To calculate the nuclear levels we use a 
relativistic shell model (see, eg. Refs.~\cite{sai96,sai07} for more details). 

\subsection{The Coulomb dissociation model}

The model used to calculate the CD cross sections of the $^{23}$O nucleus 
is described in detailed in Refs.~\cite{cha00,pra02,pra08}. Therefore, we 
give here only a brief sketch of it. This theory is formulated within the 
post-form finite range distorted wave Born approximation (FRDWBA), where the 
electromagnetic interaction between the fragments and the target nucleus is 
included to all orders and the breakup contributions from the entire 
nonresonant continuum corresponding to all the multipoles and the relative 
orbital angular momenta between the fragments are taken into account
\cite{pra02}. Only the full ground state wave function of the projectile, 
of any orbital angular momentum configuration, enters in this theory as 
input. Thus, unlike most of the theoretical models described in a recent 
review of breakup theories~\cite{bay12}, this approach does not require the 
knowledge of the positions and widths of the continuum states. Hence, it is 
free from the uncertainties associated with the multipole strength 
distributions~\cite{bau03} that may exist in some of these theories. 
 
We consider a breakup reaction ($ a + A \rightarrow b + c + A $), where the
projectile $a$ breaks up into fragments $b$ (charged) and $c$ (uncharged) in
the Coulomb field of a target $A$. The differential cross section for the
relative energy distribution for this reaction is given by
\begin{eqnarray}
\frac{d\sigma}{dE_{bc}} =\int_{\Omega _{bc},\Omega_{aA}} d\Omega _{bc}
d\Omega_{aA} \left\{\sum_{l m}\frac{1}{(2l + 1)}\vert \beta_{lm}\vert^2
\right\} \frac{2\pi}{\hbar v_{aA}} \frac{\mu_{bc}\mu_{aA}p_{bc}p_{aA}}
{h^6}~ \label{cs},
\end{eqnarray}
where $v_{aA}$ is the $a$--$A$ relative velocity in the entrance channel,
$\Omega _{bc}$ and $\Omega_{aA}$ are solid angles, $\mu_{bc}$ and
$\mu_{aA}$ are reduced masses, and $p_{bc}$ and $p_{aA}$ are appropriate
linear momenta corresponding to the $b$--$c$ and $a$--$A$ systems,
respectively.

The reduced amplitude, $\beta_{lm}$, in the post-form FRDWBA is given by
\begin{eqnarray}
\beta_{lm}
= \langle \exp(\gamma\vec{k}_c-\alpha\vec{K})\vert V_{bc}\vert
\Phi _{a}^{lm}\rangle \langle \chi_b^{(-)}(\vec{k}_b)\chi_c^{(-)}
(\delta\vec{k}_c)\vert \, \chi_a^{(+)}(\vec{k}_a) \rangle~, \label{dw}
\end{eqnarray}
where $\vec{k}_b$, $\vec{k}_c$ are Jacobi wave vectors of fragments $b$
and $c$, respectively, in the final channel of the reaction, $\vec{k}_a$
is the wave vector of projectile $a$ in the initial channel and $V_{bc}$
is the interaction between $b$ and $c$. $\Phi _{a}^{lm}$ is the ground
state wave function of the projectile with relative orbital angular
momentum state $l$ and projection $m$. In the above, $\vec{K}$ is an
effective local momentum associated with the core-target relative system,
whose direction has been taken to be the same as the direction of the
asymptotic momentum $\vec{k}_b$~\cite{cha00}. $\alpha, \delta$, and
$\gamma$ in Eq. (5), are mass factors relevant to the Jacobi coordinates
of the three body system (see Fig. 1 of Ref. \cite{cha00}). The wave
functions $\chi_b^{(-)}$ and $\chi_c^{(-)}$, are the distorted waves for the
relative motion of $b$ and $c$ with respect to $A$ and the c.m. of the
$b$-$A$ system, respectively, with ingoing wave boundary condition and
$\chi_a^{(+)}(\vec{k}_a)$ is the distorted wave for the scattering of the
c.m. of the projectile $a$ with respect to the target with outgoing
boundary conditions.

The first term in Eq.~(\ref{dw}) contains the structure information about
the projectile through the ground state wave function $\Phi _{a}^{lm}$,
while the second term is associated with the dynamics of the reaction.
For the pure Coulomb case, $\chi_b^{(-)}(\vec{k}_b)$ and $\chi_a^{(+)}
(\vec{k}_a)$ are replaced by the appropriate Coulomb distorted waves,
and $\chi_c^{(-)}(\delta\vec{k}_c)$ by a plane wave as the fragment
$c$ is uncharged (e.g. for a neutron). This  allows the second term of
Eq.~(\ref{dw}) to be evaluated analytically in terms of the bremsstrahlung
integral \cite{nor54}. A more detailed description of the evaluation of the
reduced amplitude $\beta_{lm}$  can be found in Refs.~\cite{cha00,rsh01}.
It is clear from Eqs.~(6) and (7) that within this model of breakup reactions,
explicit information about the continuum strength distribution of the projectile 
is not required in calculations of the relative energy spectra - the entire 
continuum is automatically included in this theory.

It should be remarked that this model of the Coulomb dissociation belongs to 
one particular class of breakup theories that include the interaction between 
the projectile fragments and the target nucleus to all orders but treat the 
fragment-fragment interaction in first order. Because for relative energies of our 
interest there are no resonances in the $n$+$^{22}$O continuum, we expect this 
approximation to be valid. It is clearly a good approximation for the deuteron and 
the neutron halo systems~\cite{bau03}. For those cases where higher order effects 
of the fragment-fragment interaction are known to be nonnegligible, this model will 
have a limited applicability. Recently, calculations are becoming available where 
the breakup reactions are treated in terms of the Faddeev type of theories of the 
three-body problem including also the Coulomb potentials in the fragment-target and 
fragment-fragment interactions~\cite{del09,cre09,cra10}. These studies although 
confined so far mostly to breakup reactions on a proton target, are expected to 
provide a comprehensive check on various approximations used in different types of 
breakup theories~\cite{upa12}. 

\begin{table}[t]
\begin{center}
\caption[T1]{Valence neutron separation energy as calculated in the QMC model. The second 
column shows the quantum number of the valence neutron in each case.}
\vspace{.5cm}
\begin{tabular}{|c|c|c|c|}
\hline
Isotope & valence neutron orbit & $S_{-n}^{QMC}$            & $S_{-n}^{data}$ \\
        &                       & (\footnotesize{MeV})   & (\footnotesize{MeV})
\\ \hline
$^{23}$N  & $^2s_{1/2}$ & -2.701 & -2.494 $\pm$ 0.360 \\
$^{23}$O  & $^2s_{1/2}$ & -2.858 & -2.740 $\pm$ 0.130 \\
$^{31}$Ne & $^1f_{7/2}$ & -0.004 & -0.290 $\pm$ 1.640 \\
$^{35}$Mg & $^1f_{7/2}$ & -0.861 & -0.728 $\pm$ 0.463 \\
$^{37}$Na & $^1f_{7/2}$ & -1.054 & -0.750 $\pm$ 0.180 \\
$^{45}$S  & $^2p_{3/2}$ & -1.592 & -2.208 $\pm$ 1.786 \\
$^{49}$Ar & $^2p_{3/2}$ & -2.726 & -2.501 $\pm$ 0.590 \\ 
\hline
\end{tabular}
\end{center}
\end{table}

\section{Results and discussions}

In Table 1, we show the one-neutron separation energies $(S_{-n})$  and the valence neutron 
quantum numbers of a number of light neutron rich nuclei lying in the vicinity of the 
one-neutron dripline. In the last column of this table we show the experimental value of 
$S_{-n}$ $(S_{-n}^{data})$ that have been derived from Refs.~\cite{gau12,jur07,aud03,end98}. 
It is clear that $S_{-n}$ predicted by the QMC model is in reasonable agreement with the 
corresponding experimental value for all the nuclei studied in table 1. This is remarkable 
in view of the fact that in these calculations the model parameters were the same as those 
used in the description of the normal nuclei in Ref.~\cite{gui08}. It may however, be 
remarked that for still lighter nuclei (e.g. isotopes of C and Be etc.) the QMC model may 
not be applicable as the mean field approximation is unlikely to be valid for these systems. 
For example, we found that for $^{19}$C the QMC model predicts a $S_{-n}$ of -1.192 MeV 
(-3.095 MeV) for a ground configuration in which a $2s_{1/2}$ ($1d_{5/2}$) neutron is bound 
to a $0^+$ $^{18}$O core. However, a generally accepted value of $S_{-n}^{data}$ for this 
nucleus is 0.576$\pm$0.094 MeV with a ground state spin-parity of $\frac{1}{2}^+$
\cite{nak99,shy99,des00}, even though there is still some uncertainty in the mass of 
$^{19}$C. It has been shown earlier~\cite{len01,len02} that the valence neutron in $^{19}$C 
is no longer attached to a mean field orbital. Thus, the mean-field dynamics has ceased to 
be the dominant source of binding in this case where the dynamical valence neutron-core 
interaction provides most of the binding. Therefore, a deviation of the $S_{-n}$ calculated 
in the QMC model for this nucleus from the corresponding experimental value is not surprising. 
Nevertheless, it is interesting to note that the ground state configuration $^{18}$C(0$^+$)
$\otimes$\,$n$2$s_{1/2}$ leads to a $S_{-n}$ more in agreement with its experimental value, 
which is in line with the conclusions of Refs.~\cite{nak99,shy99,des00}.

As we are particularly interested in the nucleus $^{23}$O, we discuss our results 
in some details for this case. In table 1 we notice that, the configuration where 
valence neutron lies in the $2s_{1/2}$ level (with $^{22}$O remaining as inert core), 
yields a $S_{-n}$ value of 2.86 MeV. This is very close to the experimental value 
of 2.74 MeV for this quantity. We have also tried a scheme to get a valence neutron 
spin-parity of $\frac{5}{2}^+$ by filling the last neutron orbit ($2s_{1/2}$) with 
two neutrons. This, however, makes the $^{23}$O isotope unstable in our model.  
Therefore, we confirm that the ground state of this nucleus is consistent with a 
$^{22}$O(0$^+)\otimes$2$s_{1/2}$ configuration with a one neutron separation energy 
of 2.86 MeV. It should be mentioned here that using the QMC-I model, we find a one 
neutron binding energy of 4.34 MeV within the same ground state configuration. Thus, 
the extra effects put into the QMC-III model (e.g., one-gluon exchange hyperfine 
interaction) do seem to be important to describe self-consistently the structure of 
the drip-line nuclei.  
\begin{figure}[t]
\begin{center}
\includegraphics[scale=0.50]{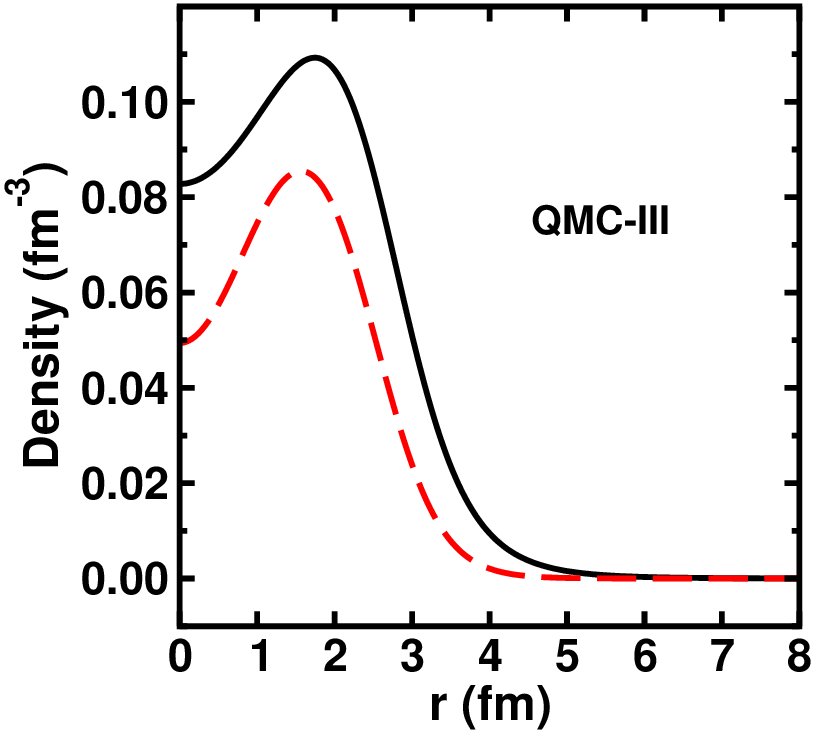}
\caption{(Color online)\label{Fig1} Neutron (solid line) and proton 
(dashed line) density distributions for $^{23}$O in QMC-III model.}
\end{center}
\end{figure}

In Fig.~1, we show the neutron and proton density distributions for $^{23}$O 
obtained within the QMC-III model. We see that the central density of the neutron 
distribution is clearly larger than that of the proton. At the same time the surface 
diffuseness of the two distributions are almost identical. This is consistent with 
the criteria for a neutron skin~\cite{cen10,jia05}. The neutron skin thickness, which 
is the  difference between the neutron and proton root mean square radii (RMS) 
\cite{miz00} is found to be 0.46 fm, which is in agreement with the value reported for 
this nucleus in Ref.~\cite{jia05} where calculations have been performed within a QHD 
type of relativistic mean field model. In neutron rich isotopes of several medium to 
heavy mass nuclei, a neutron skin of similar thickness has been reported by several 
authors (see, e.g.  Refs.~\cite{wie11}).

In Figs.~2a and 2b, we show the valence neutron-core potential and the corresponding 
wave function, respectively, for the $^{23}$O nucleus calculated within the QMC-III 
model. The potential is the sum of scalar and vector fields and the wave function is 
only the upper component of the spinor, as the corresponding lower component is at 
least two orders of magnitude smaller. For comparison, we also show a Woods-Saxon (WS) 
potential and the respective wave function for this system. This potential has been 
parameterized by adjusting its depth to reproduce the experimental value of $S_{-}n$ with 
the radius and diffuseness parameters of 1.15 fm and 0.5 fm, respectively. We note that 
while the QMC-III potential is somewhat deeper than the WS potential at smaller 
distances, they are similar in the surface region. However, the wave functions generated 
by the two potentials are almost identical at all radii. 
\begin{figure}[t]
\begin{center}
\includegraphics[scale=0.50]{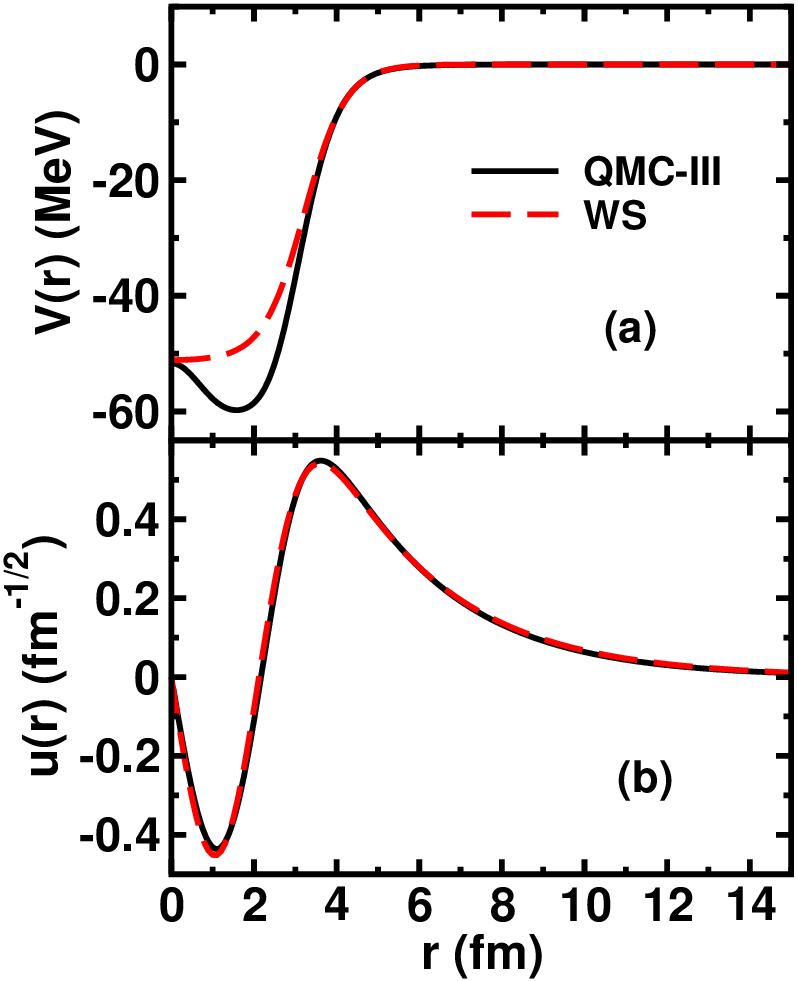}
\caption{(Color online)\label{Fig2} The QMC-III (solid line) and Woods-Saxon 
(dashed line) valence neutron-core potentials [part (a)] and corresponding 
wave functions [part (b)] for the $^{23}$O nucleus.}  
\end{center}
\end{figure}

The wave functions calculated above have been used to investigate the Coulomb 
dissociation (CD) of the $^{23}$O nucleus on Pb target, for which some data exists
\cite{noc05} on the valence neutron-core relative energy differential cross section and 
the total one-neutron removal cross sections. The CD process has a number of advantages. 
It is free from the uncertainties associated with the nuclear interactions. Moreover the 
inelastic breakup mode (also known as stripping or breakup fusion) is absent in the pure 
Coulomb dissociation reactions. Therefore, this process is ideally suited for probing 
the structure of the projectile.

In Fig.~3, the calculated energy differential cross section ($\frac{d\sigma}{dE_x}$) 
is shown as a function of the excitation energy ($E_x$ =  $E_{bc}$ + $S_{-n}$, where 
$E_{bc}$ is the n-$^{22}$O relative energy). The results obtained with QMC-III 
(solid line) as well as a WS ground state (dashed line) wave functions are displayed. 
In getting these cross sections, an integration has been carried out over 
$\theta_{aA}$ up to the grazing angle of the reaction. In order to compare the 
calculations with the data of Ref.~\cite{noc05}, the calculated cross sections need 
to be convoluted with the detector response function of the experiment. Since this is 
not available to us we have normalized our cross sections to the peak of the 
experimental $\frac{d\sigma}{dE_x}$ in each case. This may be viewed as an alternative 
to the convolution procedure.
\begin{figure}[t]
\begin{center}
\includegraphics[scale=0.50]{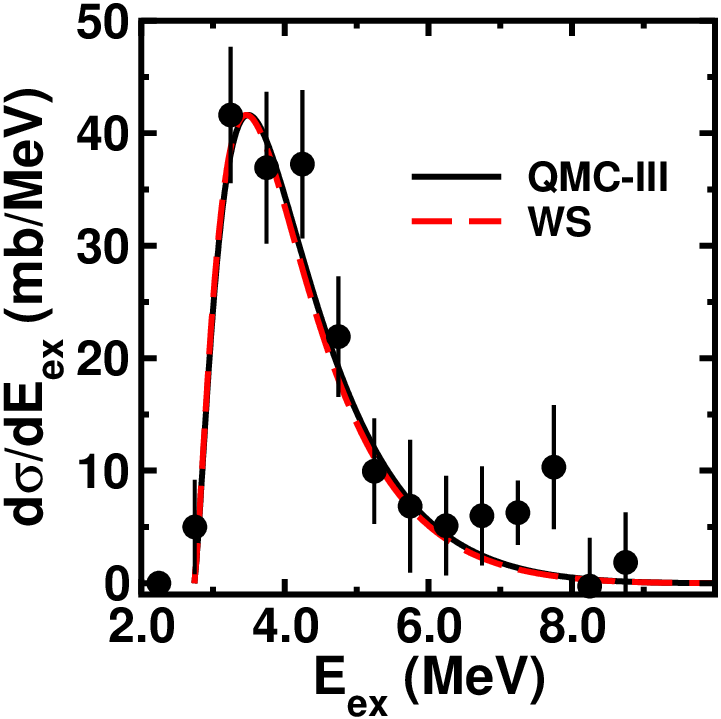}
\caption{(Color online)\label{Fig3} The calculated excitation energy spectra
in the Coulomb dissociation of $^{23}$O on a Pb target at 422 MeV/nucleon 
beam energy obtained with QMC-III (solid line) and Woods-Saxon (dashed line) 
wave function. The data is taken from Ref.~\protect\cite{noc05}. The 
calculated cross sections are normalized to the peak of the experimental 
data in each case.} 
\end{center}
\end{figure}

We note that our calculations are able to describe the shape of the relative energy 
spectrum quite well over the entire region of excitation energies. The sharp rise of 
the experimental cross section just after the threshold is quite well reproduced. 
Even the widths of the experimental distribution is well reproduced in our model. In 
contrast to this, the width of the distribution is grossly over predicted within a 
semiclassical breakup model using plane waves for the relative motion wave functions 
in the outgoing channel, as shown in Ref.~\cite{noc05}. Once optical potentials are 
used to describe the n $ + ^{22}$O relative motion, the situation improves. , 

The total electromagnetic one-neutron removal cross sections obtained in 
our CD model is 79.26 mb and 77.15 mb with QMC-III and WS wave functions, 
respectively. This is only about 10$\%$ smaller than the lower limit of the 
corresponding experimental value of 97$\pm$10 mb~\cite{noc05}. However,
since the upper limit of uncertainty in the experimental data is of the 
order of 20$\%$~\cite{noc05}, this slight under prediction of the 
total electromagnetic cross sections by our model is not significant. 

In Fig.~4, we show the calculated LMD for the $^{22}$O fragment emitted in 
the CD of $^{23}$O on a Pb target at 422 MeV/C beam energy, calculated 
with QMC-III and WS ground state wave functions. The 
full width at half maximum (FWHM) of the LMD is about 85 MeV/c in both  
cases. This value is more than a factor 2 larger than the FWHM of the 
LMD of the core fragment observed in reactions induced by established 
one-neutron halo nuclei like $^{11}$Be and $^{19}$C.  Although there are 
no data available for the LMD of the $^{22}$O fragment in the $^{23}$O 
induced reaction on a heavier target like Pb, the FWHM of the LMD of 
$^{22}$O have been measured in the $^{23}$O induced breakup reaction on 
light targets like carbon~\cite{kan02,gil04,taj10}, where they have been 
found to lie around 100 MeV/c. Therefore, these results almost rule out 
the development of a one-neutron halo structure in $^{23}$O, even though 
the valence neutron occupies an s-orbit in this nucleus.
\begin{figure}[t]
\begin{center}
\includegraphics[scale=0.50]{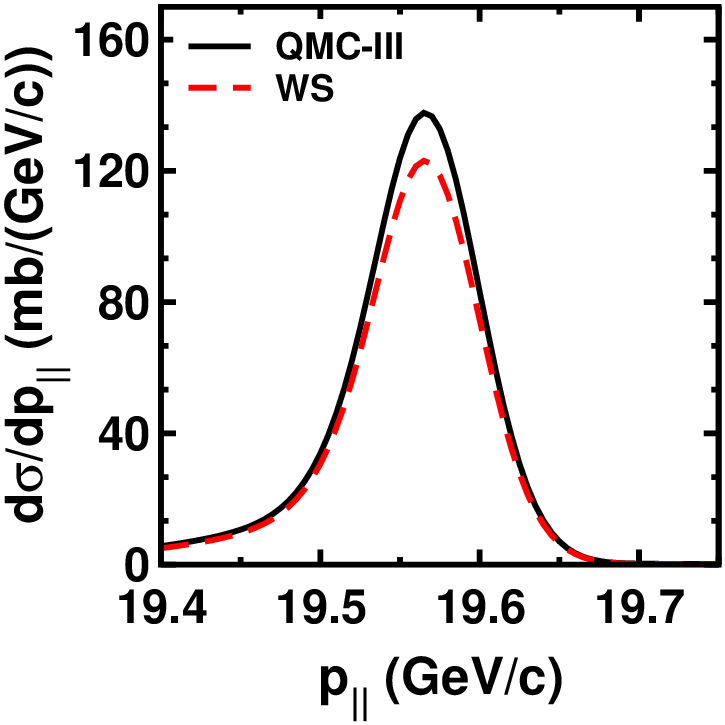}
\caption{(Color online)\label{Fig4} The calculated longitudinal momentum 
distribution of $^{22}$O fragment in the Coulomb dissociation of $^{23}$O on 
a Pb target at 422 MeV/nucleon beam energy obtained with QMC-III (solid line) 
and Woods-Saxon (dotted line) wave function.
} 
\end{center}
\end{figure}

\section{Summary and conclusions}

In summary, we have studied the ground state structure and the one neutron
separation energies of a number of light mass neutron rich nuclei lying in the 
vicinity of one-neutron drip line within the latest version of the quark-meson 
coupling model. With the same set of parameters that were used to describe the 
properties of normal nuclei, we obtain a reasonable agreement with the experimental
one-neutron separation energies. We concentrated particularly on the $^{23}$O
nucleus. Our calculations confirm a $^{22}$O(0$^+)\otimes$2$s_{1/2}$ configuration 
for the ground state of this nucleus with a one neutron separation energy
of 2.86 MeV that is in close agreement with the corresponding experimental value. 
A spin-parity assignment of $1d_{5/2}$ to this nucleus is not supported by 
our model. We predict the development of the neutron skin in this nucleus 
with a thickness of 0.48 fm.

The valence neutron-core wave function calculated within the QMC model has been 
used to study the Coulomb dissociation of $^{23}$O on a Pb target at the beam energy 
of 422 MeV/c employing a theory that requires only the ground state projectile wave 
function as input and is free from any other adjustable parameter. Although at this 
beam energy the relativistic effects play a role~\cite{ber05}, yet a fully quantal 
relativistic theory of breakup reactions is not yet available. Our theory is 
essentially non-relativistic in nature. Nevertheless, we observe that the existing 
data on the excitation energy spectra and the total electromagnetic one-neutron 
removal cross section are well reproduced.  The calculated longitudinal momentum 
distributions of the $^{22}$O fragment are broad, which effectively excludes 
the development of a one-neutron halo structure in this nucleus even though the valence 
neutron occupies an $s$-orbit. Our study shows that the latest quark-meson coupling 
model provides a competitive alternative for describing the structure of the drip 
line nuclei as compared to the commonly used non-relativistic models. 

\section{Acknowledgments}

This work has been supported by the University of Adelaide and the 
Australian Research Council through grant FL0992247(AWT) and through the 
ARC Centre of Excellence for Particle Physics at the Terascale.  KT was 
also supported by a visiting professorship of International Institute of 
Physics, Federal University of Rio Grande do Norte, Natal, Brazil.

\end{document}